\documentclass[showpacs,aps,prb,twocolumn,footinbib]{revtex4} 

\usepackage{graphicx,amssymb,amsmath} 
\usepackage{dcolumn}
\usepackage{bm}

\begin{document}

\title{Diffusion-limited exciton-exciton annihilation in single-walled carbon nanotubes: \\ A time-dependent analysis}

\author{Ajit Srivastava}
\thanks{corresponding author}


\author{Junichiro Kono}
\affiliation{Department of Electrical and Computer Engineering, Rice University, Houston, Texas 77005, USA}

\date{\today}

\begin{abstract}
To provide physical insight into the recently observed photoluminescence saturation behaviors in single-walled carbon nanotubes implying the existence of an upper limit of exciton densities, we have performed a time-dependent theoretical study of diffusion-limited exciton-exciton annihilation in the general context of reaction-diffusion processes, for which exact treatments exist.
By including the radiative recombination decay as a Poissonian process in the exactly-solvable problem of one-dimensional diffusion-driven two-particle annihilation, we were able to correctly model the dynamics of excitons as a function of time with different initial densities, which in turn allowed us to reproduce the experimentally observed photoluminescence saturation behavior at high exciton densities.  We also performed Monte Carlo simulations of the purely stochastic, Brownian diffusive motion of one-dimensional excitons, which validated our analytical results.  Finally, we consider the temperature-dependence of this diffusion-limited exciton-exciton annihilation and point out that high excitonic densities in SWNTs could be achieved at low temperature in an external magnetic field.

\end{abstract}


\maketitle

\section{Introduction}
Excitons in single-walled carbon nanotubes (SWNTs) are stable quasi-particles with large binding energies and significantly influence their interband optical properties.\cite{Dresselhaus08ExcitonReview}  However, they have been reported to be rather efficiently eliminated at high densities through the exciton-exciton annihilation (EEA) process,\cite{Ma-PRL-2005} although their emission and absorption energies remain stable even at high densities.\cite{Ostojic-PRL-2005}  Recently, the intensity of photoluminescence (PL) from SWNTs was found to saturate at high pump fluence,\cite{Murakami-under-review} implying the existence of an upper limit in the density of excitons, which was estimated to be an order of magnitude smaller than the expected Mott density.  The existence of such an upper limit, which poses a significant hinderance in the observation of lasing, a Mott transition, or excitonic Bose-Einstein condensation in SWNTs, was attributed to efficient EEA facilitated by the diffusive motion of excitons.\cite{Sheng-PRB-2005,Russo-PRB-2006,Cognet-Science-2007}

The dynamics of diffusion-limited EEA can be analyzed in the general context of reaction-diffusion processes, which have been extensively studied by physicists, chemists, biologists, and ecologists and serve as simple models for studying a variety of non-equilibrium problems.\cite{BookBenAvraham,CardyMathBeauty,BookNonEqlbTransMarro,Odor04RMP,BookOkubo} Moreover, it has been shown that such simple diffusion-driven reactions exhibit interesting non-equilibrium phase transitions and universality classes,\cite{Hinrichsen00AdvPhys} with connections to many-body theory.\cite{KamanevManybody1,BookKamenevManybody2}
In such models, particles, or ``agents," of one or more species execute random walk in $d$-dimensions (where $d$ could also be fractional in the case of fractal geometry) and undergo reactions upon collisions, leading to changes in their population often accompanied by an appearance and disappearance of various phases.  Such systems often exhibit rich phase diagrams that can be fully studied with numerical simulations even when analytical solutions are not available.  For example, a widely studied reaction is the two-particle annihilation given by $A + A$ $\rightarrow$ 0, where the interaction is assumed to be of the ``hard-core" type, leading to the mutual destruction of two particles upon collision.\cite{CardyMathBeauty}  Starting with an initial population $N_0$, one can analyze the ensemble averaged population $N_t$ at a given time $t$.  Clearly, this is a non-equilibrium many-particle process that is driven by noise, and its only steady state is achieved when the population vanishes.

An interesting feature of diffusion-driven reactions is the presence of spatio-temporal fluctuations in such processes that cannot be ignored especially at lower dimensions and lead to the breakdown of mean-field type assumptions.  For example, diffusion-driven two-particle annihilation can be written in differential form as follows:
\begin{equation}
\partial_{t}\langle n(x,t)\rangle = D\nabla^{2}\langle n(x,t)\rangle - \langle n^{2}(x,t)\rangle,
\label{differentialform}
\end{equation}
where $\langle . \rangle$ stands for an ensemble average.  The first term on the right denotes the diffusion process, while the second term is for two-particle annihilation and thus has quadratic dependence.  Note that the annihilation term has an average of $n^{2}(x,t)$ whereas mean-field theory will simplify this to $\langle n(x,t)\rangle^{2}$, thus neglecting fluctuations of the form $\langle n^{2}(x,t)\rangle-\langle n(x,t)\rangle^{2}$.  In one dimension, the population asymptotically decays as $t^{-1/2}$ power-law whereas the mean-field theory, which ignores fluctuations, predicts a faster decay of $t^{-1}$.  This discrepancy originates from the fact that $d$-dimensional diffusion with $d \leq 2$ is \emph{recurrent} and the particles return to their previous position with high probability.\cite{PolyaResult}  Hence, the reaction is slowed down leading to a smaller exponent in the power-law decay.  Indeed, the mean-field result is recovered in three dimensions, which is above the critical dimension $d_{c}$ = 2 for this problem.\cite{CardyTauber98JStatPhys}  Thus, an exact treatment of even such a simple process requires the inclusion of correlations.


In this paper, we undertake a time-dependent study of the diffusion-limited EEA process in the presence of radiative decay.  In particular, we consider the following two coupled and competing reaction-diffusion processes in 1-D to model the dynamics of excitons in nanotubes at various densities:
\begin{subequations}
\begin{align}
A + A &\rightarrow kA \quad (k = 0, 1);\\
A &\xrightarrow{\gamma_{r}} B
\end{align}
\label{RD1}
\end{subequations}
where $A$ represents excitons and $B$ photons.  The first equation represents exciton-exciton annihilation, which is either complete ($k$ = 0) or partial ($k$ = 1), while the second reaction is just the radiative decay of excitons with radiative lifetime $\tau_r$ = $1/\gamma_{r}$.  It is noteworthy that only the first reaction is diffusion-driven whereas as the radiative decay takes place independently.  This diffusive motion of excitons is due to the random collisions with phonons.\cite{Cognet-Science-2007}  In this sense, the annihilation reaction is driven by diffusive noise, which we assume to be of the Gaussian form, whereas the radiative decay is governed by a Poissonian noise and, hence, is a pure jump process.  We consider a simple diffusion process in which the diffusion constant $D$ is independent of the spatial and temporal coordinates.  Furthermore, the excitonic dimension is assumed to be much smaller than the nanotube dimension.  We are interested in determining the population of both species as a function of time.  The population of species $B$, or photons, is proportional to the PL intensity measured experimentally.  In particular, we wish to know the fraction of population which decays radiatively and how this fraction changes as the initial population is increased.  As one can imagine, upon increasing the initial density of excitons in the 1-D nanotube the annihilation reaction becomes more efficient whereas the radiative decay rate can be safely assumed to be independent of the density.  As we show later, this leads to saturation of the PL intensity as the initial population density is increased, which is consistent with the experimental observations.\cite{Murakami-under-review}

\section{Solution using First-Passage distribution}

The case of two-particle annihilation without the radiative decay has been extensively studied, and exact results for the population as a function of time are known.\cite{Torney-JPC-1983,Toussaint-JCP-1983,ben-Avraham-PRL-1998,Spouge88PRL}  Here we use the first-passage time distribution of Brownian motion to first study the annihilation reaction without decay.  We recover the exact analytical results for this case before proceeding to include the radiative decay term.  We begin by deriving the exact result for a single pair, or the ``independent pairs" case, and use it to obtain an approximate solution for the many-particle, or the ``correlated" case.  Monte Carlo simulations are performed to check the validity of our results.  This purely stochastic method employing the first-passage time distribution is a simple and natural way to study the annihilation reaction as the collisions which drive the reactions must obey such distributions.

\subsection{Annihilation without decay; $k$ = 0}
Consider $N_0$ pairs of species $A$ randomly arranged on a line of length $L$ and executing Brownian diffusion with diffusion constant $D$.  Let us first consider just the two-particle annihilation process without the decay as in Eq.~(\ref{RD1}a) with $k$ = 0.  Let $n_{A}(t)$ denote the average fraction of initial population which is still ``alive" at time $t$.  We keep the discussion in this section as general as possible without explicitly identifying species $A$ or $B$ unless absolutely required.

In 1-D, only the nearest neighbors at any given time can undergo annihilation due to restrictions placed by lower dimensionality.  This prompts us to first consider the case for a single pair of particles and generalize the result to the many-particle case.  As this is equivalent to different pairs annihilating independently of each other, we refer to it as the ``independent pair" case.  Let $d_{0}$ be the initial distance between the pair and $P(d_{0},t)$ denote the survival probability of this pair at time $t$.  To calculate $P(d_{0},t)$, we need to find the probability that a pair with initial distance $d_{0}$ does not undergo collision till time $t$.  As both particles are executing independent Brownian motion, their relative motion is also Brownian with diffusion constant $2D$ which starts at $d_{0}$.  Thus, we need to find the probability that a Brownian motion starting at $d_{0}$ does not reach zero till time $t$.  This can be readily found from the first-passage time distribution of a Brownian motion as\cite{BookShreve}
\begin{equation}
P\left(d_{0},t\right) = \mathrm{erf}\left(\frac{d_{0}}{2\sqrt{Dt}}\right),
\label{survival probability}
\end{equation}
where $\mathrm{erf}(.)$ is the error function.  For the independent pair case, the fraction of population that is alive at time $t$, $n_{A}(t)$, reads
\begin{equation}
n_{A}\left(t\right)=\sum_{i=1}^{N_0}\mathrm{erf}\left(\frac{d_{0,2i}}{2\sqrt{Dt}}\right)=\sum_{i=1}^{N_0}\mathrm{erf}\left(\frac{d_{0,2i-1}}{2\sqrt{Dt}}\right),
\label{summationform}
\end{equation}
where the distance between the particles of the $i$-th pair is $d_{0,i}$ at time $t=0$.  We have imposed a periodic boundary condition making the line into a ring without any loss of generality.  As $N_0$ tends to infinity, the above sum can be expressed as an integral over the distribution of $d_{0,i}$s.  We restrict ourselves to the case when the particles are randomly arranged on the line at $t$ = 0.  Thus, $d_{0,i}$s which are the nearest neighbor distances can be thought of as the ``waiting times" for a Poisson process and have an exponential distribution with mean $d_0$ = $L/2N_{0}$.

For the many-particle case we realize that there are twice as many ways for a pair to annihilate as for the independent case due to the presence of two nearest neighbors for each particle, and hence, the mean distance for the correlated case is just a half of the independent case in the large $N_{0}$ limit.  For this limit, the exact result can be obtained as
%
\begin{align}
n_{A}\left(t\right) &= \int_{0}^{\infty}dx\beta\mathrm{exp}\left(-\beta x\right)\mathrm{erf}\left(\frac{x}{2\sqrt{Dt}}\right) \nonumber \\
&= \mathrm{exp}\left(\beta^{2}Dt\right)\mathrm{erfc}\left(\sqrt{\beta^{2}Dt}\right)
\label{largeNanalnodecay}
\end{align}
%
which is the result of Torney {\it et al}.\cite{Torney-JPC-1983} with $\beta$ = $4N_{0}/L$.  In the asymptotic limit, Eq.~(\ref{largeNanalnodecay}) yields a power-law decay of $t^{-1/2}$ as mentioned earlier.

At long times the initial correlations between the particles are completely wiped out, and this power-law decay is expected, irrespective of the initial distribution of particles.  Such a power-law behavior has been recently observed in time-resolved transient absorption measurements on SWNTs.\cite{Russo-PRB-2006}

\subsection{Annihilation with decay; $k$ = 0: Independent pair case}
Next, we include radiative decay of Eq.~(\ref{RD1}b) in our model and compute the population fraction of species $A$ and $B$ as a function of time.  As before, we first derive the exact result for the case of a single pair and use it as a kernel to express the result for independent pairs uniformly distributed along the tube.  For a single pair separated by a distance $d_{0}$ at $t=0$, the surviving population fraction at $t$ can be simply written as
\begin{equation}
n_{A}\left(t\right)=\frac{1}{2}\sum n p(n,t) \quad (n=0,1,2)
\label{singlepair1}
\end{equation}
where $p(n,t)$ denotes the probability of $n$ surviving particles at time $t$, which remains to be calculated.  Let us compute $p(2,t)$, which is the probability that both particles comprising the pair are alive at time $t$.  Such a case is possible \emph{only} if neither particle undergoes radiative decay or collision till time $t$.  As the radiative decay of particles occurs independently of one another and also of the diffusion driven collision, we can simply multiply the individual probabilities to get
\begin{equation}
p(2,t)=\mathrm{exp}\left(-2\gamma_{r} t\right)\mathrm{erf}\left(\frac{d_{0}}{2\sqrt{Dt}}\right).
\label{p(2,t)}
\end{equation}
Recall that radiative decay is a Poisson process with parameter $\gamma_{r}$ and the probability of it not happening till time $t$ is $\mathrm{exp}\left(-\gamma_{r} t\right)$.  To compute $p(1,t)$, which is the probability that \emph{exactly} one particle out of the pair survives, we realize that such a scenario is possible \emph{only} if there is exactly one radiative decay in the time interval $[0,t]$, say at $t=\tau$, and no collision \emph{before} $\tau$.  As in the interval $(\tau, t]$ collisions cannot take place due to an insufficient number of particles for the reaction, we only include the probability of collision not taking place before $\tau$.  The probability that either of the particles decay in an infinitesimal interval $d\tau$ about $\tau$ is $2\gamma_{r} d\tau$.  As before, we can multiply the probabilities for each sub-event due to mutual independence.  Thus,
\begin{equation}
\begin{split}
p(1,t)=&\int_{0}^{t}\mathrm{exp}\left(-2\gamma_{r} \tau \right)\mathrm{erf}\left(\frac{d_{0}}{2\sqrt{D\tau}}\right)\\
&\times \left(2\gamma_{r} d\tau\right) \mathrm{exp}\left(-2\gamma_{r}\left(t-\tau\right)\right).
\end{split}
\label{p(1,t)}
\end{equation}
From Eqs.~(\ref{singlepair1}), (\ref{p(2,t)}), and (\ref{p(1,t)}), $n_{A}(t)$ for a single pair can be obtained.  As before, for the case of $N_0$ independent pairs, we average $n_{A}(t)$ over an exponential distribution with mean $d_0$ = $L/2N_{0}$.  After some straightforward but tedious algebra we obtain
\begin{equation}
\begin{split}
n_{A}(t)=&\frac{\mathrm{exp}\left(-\gamma_{r} t \right)}{1-\nu}\left[\mathrm{exp}\left( \frac{1-\nu}{\nu}\gamma_{r} t\right)\mathrm{erfc}\left(\sqrt{\gamma_{r} t/ \nu}\right)\right. \\
&\left.+\sqrt{\nu}\mathrm{erf}\left(\sqrt{\gamma_{r} t}\right)-\nu\right],
\label{exactindeppair}
\end{split}
\end{equation}
where we have introduced a dimensionless parameter $\nu$ = $\tau_{D}/\tau_{r}$ with $\tau_{D}$ being the ``diffusional time'' $d_{0}^{2}/D$.  We emphasize that this is an {\em exact} result for the case of independently colliding pairs that also undergo radiative decay.  The above equation is valid only when $\nu$ $<$ 1, or, in other words, when radiative decay is slower.  In the limit of extremely dilute initial population density, no annihilation can take place, and only radiative decay occurs.  $n_{A}(t)$ would be simply $\mathrm{exp}\left(-\gamma_{r} t\right)$ in that case.  In the opposite limit when radiative decay rate $\gamma_{r}$ vanishes, Eq.~(\ref{exactindeppair}) indeed recovers the result of Eq.~(\ref{largeNanalnodecay}), as expected.

\begin{figure}
\includegraphics[scale=0.51]{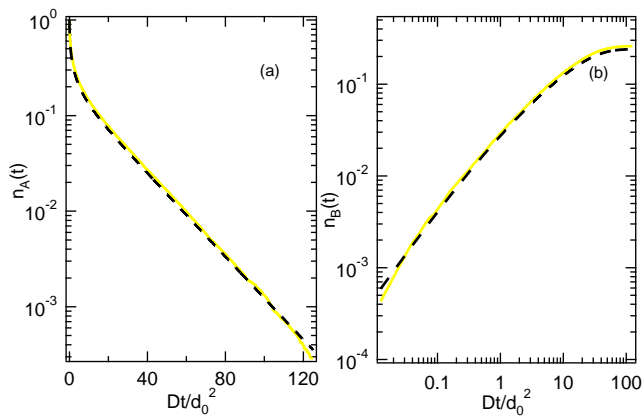}
\caption{(color online).  Comparison of Monte Carlo simulations (solid line) and exact analytical result (dashed line) for the $k$ = 0, ``independent" case with radiative decay [Eq.~(\ref{exactindeppair})]. The fraction of $A$ [in (a)] and $B$ [in (b)] populations is plotted as a function of dimensionless time $Dt/d_{0}^{2}$.  The value of $D$ = 100, $\tau_{r}$ = 80 and $d_{0}$ = 20 ($\nu$ = 0.05) were used for illustrative purposes.
}
\label{indepsimulcomp}
\end{figure}

Figure \ref{indepsimulcomp} compares the result of Eq.~(\ref{exactindeppair}) with Monte Carlo simulations, done by simulating the Brownian diffusive motion of each independent pair of particles, validating our results.  As $\nu$ is the only physical parameter in the problem, changing $d_{0}$ and $D$ but keeping $\nu$ constant should not alter the result, which was indeed confirmed by simulations. This fact should remain true even for the case of correlated pairs.

Let us consider the behavior of $n_{A}(t)$ in the long and short time limits.  For $t$ $\gg$ 1, only the radiative decay should dominate as the density of particle becomes too low to participate in annihilation.  Thus, an exponential decay is expected.  Taking limits explicitly, one obtains
\begin{equation}
n_{A}(t\rightarrow \infty) = \frac{\mathrm{exp}\left(-\gamma_{r}\right)}{1+1/\sqrt{\nu}}.
\label{n_A_asymp}
\end{equation}
As $\nu$ is less than unity, so is the intercept of the above exponential decay, hinting at the superexponential decay at short times due to annihilation.  In the short time limit, when $\nu$ $<$ 1 one expects only annihilation to dominate the decay, and one gets
\begin{equation}
n_{A}(t\rightarrow 0) = 1-2\sqrt{\gamma_{r} t/\pi\nu}.
\label{n_A_0}
\end{equation}
which is indeed faster than an exponential decay as $t$ $\rightarrow$ 0.  Indeed, Fig.~\ref{indepsimulcomp} confirms these findings. As the time progresses, the density of particles decreases due to decreasing population, which slows down the annihilation reaction as it strongly depends on the density of the particles.  The radiative decay rate, on the other hand, is fixed, and thus, a crossover from annihilation dominated decay to a purely exponential radiative decay is expected.  It can be defined to take place when $d_{t}$ = $L/2N_{0}n_{A}(t)$ becomes equal to $\nu$.  This time $\tau^{*}$ is implicitly given as
\begin{equation}
n_{A}(\tau^{*}) = \sqrt{\nu}.
\label{crossover}
\end{equation}
As the initial density is increased, $\tau^{*}$ decreases finally vanishes at very high density, implying a purely exponential decay at all times.

In order to calculate $n_{B}(t)$, we note that at any given time the rate of radiative decay is proportional to the instantaneous population of $A$, $n_{A}(t)$. In other words,
\begin{equation}
\partial_{t}n_{B}(t) = \gamma_{r} n_{A}(t).
\label{PDE_n_B}
\end{equation}
Hence, $n_{B}(t)$ can be obtained from Eq.~(\ref{exactindeppair}), upon direct integration, as
\begin{equation}
n_{B}(t)=\gamma_{r}\int_{0}^{t}d\tau n_{A}(\tau).
\label{n_B(t)}
\end{equation}
In particular, the fraction of total population that decays radiatively is given by
\begin{equation}
n_{B}(\infty)=\frac{1}{1+1/\sqrt{2\nu}}
\label{n_B(inf)}
\end{equation}
%
\begin{figure}
\includegraphics[scale=0.58]{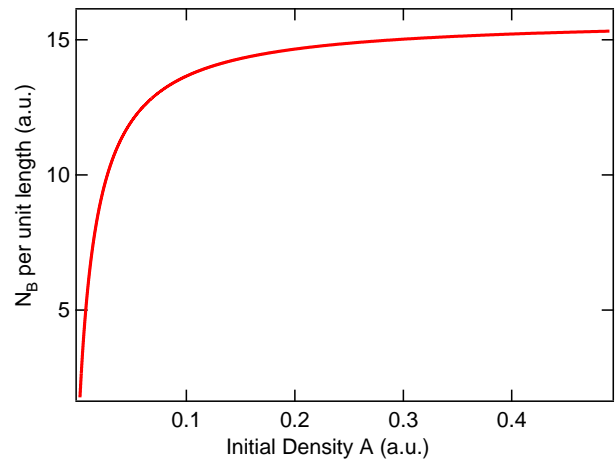}
\caption{
(color online).
Population per unit length of $B$ plotted against the initial density of $A$ using Eq.~(\ref{n_B(inf)}) to show the saturation behavior in PL intensity predicted by our model.
}
\label{Fig_n_B(inf)}
\end{figure}
In Fig.~\ref{Fig_n_B(inf)}, we use Eq.~(\ref{n_B(inf)}) to plot the total number density of species $B$, or photons, created as a function of initial density of species $A$, or excitons.  A saturation behavior of the number of photons created as the density of excitons is increased is seen even for the independent pair model agreeing with our intuitive understanding and experimental observations.  Thus, the independent pair model captures all the essential features of the process.  More importantly, the study of independent pair model identifies the relevant parameters of the process and the scaling relationships they must obey. It also provides the decay regimes that are relevant for each type of reaction viz., annihilation and radiative decay.  The use of purely stochastic first-passage time distribution makes the solution transparent and simple, relying on the properties of diffusion rather than other formal methods, which although more general are less intuitive.

\subsection{Annihilation with decay; $k$ = 0: Correlated pair case}
\begin{figure}
\includegraphics[scale=0.52]{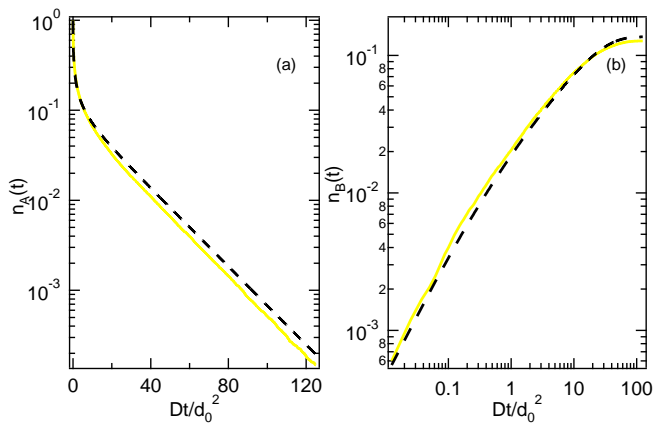}
\caption{
(color online).
Comparison of Monte Carlo simulations (solid line) and approximate analytical result (dashed line) for $k$ = 0 ``correlated pair" case with radiative decay (Eq.~\ref{exactindeppair}).  (a) The population fraction of $A$ (a) and $B$ (b) as a function of ``dimensionless" time.  Approximate result agrees fairly well with the simulations.  The parameters for simulations are the same as in Fig.~\ref{indepsimulcomp} except with $d_{0}$ = 10 ($\nu$ = 0.025).  Some reasons for disagreement with the simulations are discussed in the text.
}
\label{correlatedsimul}
\end{figure}

As for the case of no decay, we scale the mean separation between the particles by a factor of two in order to obtain a solution for the correlated case.  This approximate solution and the Monte Carlo simulations for the correlated case are compared in Fig.~\ref{correlatedsimul}.  The approximate solution agrees well with the simulations. A possible reason for the slower decay of analytical result compared to the exact result could be the following:  For the single pair case, if one of the particle decays before undergoing collision, the remaining particle \emph{must} decay radiatively and cannot undergo annihilation.  However, for the correlated case, this is not true as long as there are other neighboring particles and so annihilation becomes possible.

\subsection{Annihilation with/without decay; $k$ = 1}
\begin{figure}
\includegraphics[scale=0.51]{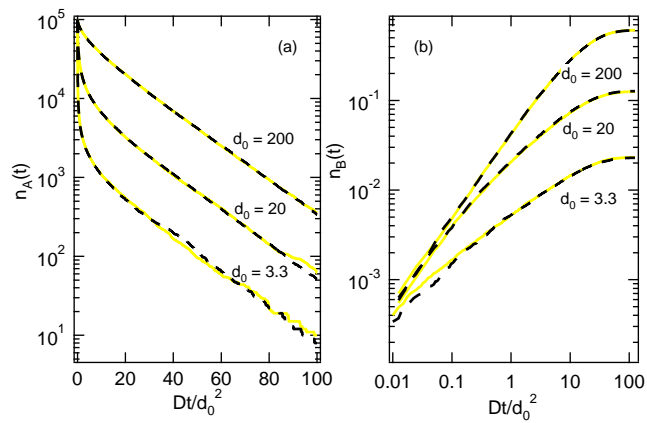}
\caption{
(color online).
Comparison of $k$ = 1 (solid line) and $k$ = 0 (dashed line) for different values of $d_{0}$ for the ``correlated" case with radiative decay.  (a) The population fraction of $A$ (a) and $B$ (b) as a function of ``dimensionless" time.  $d_{0}$ values for $k$ = 1 case simulations are half of the $k$ = 0 case shown on the graph).  Rest of the parameters are the same as in Fig.~\ref{indepsimulcomp}
}
\label{k0k1comparison}
\end{figure}

The case of partial annihilation [$k=1$ in Eq.~(\ref{RD1}a)] can be understood in terms of the results for $k$ = 0.  Both processes are completely identical besides the fact that the annihilation in $k$ = 1 is half as slow as the $k$ = 0 case.  Consequently, if the initial density for the partial annihilation case is twice as much as the complete annihilation case, one expects the two decays to be identical.  Thus, the result for the $k$ = 1 case can be obtained from Eq.~(\ref{largeNanalnodecay}) by replacing $d_{0}$ with $d_{0}/2$.  Even in the presence of radiative decay, the above argument should be true as the radiative decay occurs completely independently of the annihilation reaction.  We verify this heuristic reasoning by Monte Carlo simulations, as shown in Fig.~\ref{k0k1comparison}.

\section{Effect of temperature on the upper limit of exciton density}

Thus far, we have not included the effect of temperature ($T$) in our discussions.  The temperature-dependence of the diffusion constant $D$ can be approximated through the Einstein relation,
\begin{equation}
D=\frac{k_{B}T}{M\Gamma},
\label{EinsteinEqn}
\end{equation}
where $M$ is the mass of the exciton and $\Gamma$ is the exciton-phonon scattering rate.  At low temperatures, acoustic phonon scattering is expected to be dominant, and $\Gamma$ for 1-D is then given in terms of the deformation potential $D_{dp}$ as
\begin{equation}
\Gamma = \frac{\sqrt{2M}D_{dp}^{2}\sqrt{k_{B}T}}{\hbar^{2}\rho v_{s}^{2}},
\label{acousticphononscatteringrate1D}
\end{equation}
where $\rho$ is the mass density and $v_{s}$ is the sound velocity in the material.  Thus, the temperature dependence of $D$ due to acoustic phonon scattering is $D \propto \sqrt{T}/M^{3/2}$.  As the temperature is increased, $\Gamma$ would become a sum of both acoustic and optical phonon scattering rates.

As the relevant quantity in our model is $\nu$ and not just $D$, we need to include the temperature dependence of $\tau_{r}$ in order to fully understand the temperature dependence of the EEA process in the presence of radiative recombination.  The radiative lifetime $\tau_{r}$ of a \emph{single} 1-D exicton band is predicted to scale as\cite{Citrin92PRL} $\sqrt{T}$, leading to $\nu$ $\propto$ $1/T$ or the exciton diffusion length $l_{X}$ $\propto$ $\sqrt{T}$.  Thus, the saturation of photoluminescence due to EEA would become less effective at lower temperatures.  However, in the case of carbon nanotubes, the presence of optically inactive, or ``dark,'' states lying below the optically active, or ``bright,'' state causes radiative lifetime to increase at low temperatures,\cite{SpataruetAl05PRL,PerebeinosetAl05NL} and this could favor the EEA process depending on the exact temperature dependence.  By applying symmetry breaking perturbations such as a magnetic field, the dark state can be brightened,\cite{SrivastavaetAl08PRL} restoring the temperature dependence for the case of a single 1-D exciton band at higher fields.  Under such conditions, it may be possible to attain the Mott density of excitons in carbon nanotubes.  Sustaining such high densities of excitons is the first step for any lasing applications and for observing excitonic Bose-Einstein condensation in carbon nanotubes.  In addition, at lower temperatures exciton localization due to impurity traps or defects could completely stop the diffusive motion of excitons, further enabling the attainment of the Mott density.\cite{HogeleetAl07PRL}

Finally, in our model, we have assumed a completely random motion of excitons, which leads to ordinary diffusion based on a Gaussian kernel.  This assumption can also break down at lower temperatures or in other scenarios when $D$ becomes position- or density-dependent, leading to anomalous diffusion and changing the time-dependence of excitonic population.  A time-resolved experiment, probing the excitonic or the photon population at different temperatures, exciton densities, and magnetic fields, can not only verify the validity of this model but also provide further insight into the EEA process in carbon nanotubes.

\section{Summary}

To provide fundamental physical insights into the recently observed photoluminescence saturation behaviors in single-walled carbon nanotubes, we studied the diffusion and two-particle annihilation of one-dimensional excitons in the general context of reaction-diffusion processes, for which exact treatments exist.
By including the radiative recombination decay as a Poissonian process in the exactly-solvable problem of one-dimensional diffusion-driven two-particle annihilation, we were able to correctly simulate the density of excitons in single-walled carbon nanotubes as a function of time and density.  Monte Carlo simulations were also performed by simulating the purely stochastic, Brownian diffusive motion of one-dimensional excitons, validating our results.  Finally, we discussed the temperature dependence of EEA and proposed possible experiments to verify the validity of this model.

\begin{acknowledgments}
We thank the Robert A.~Welch Foundation (Grant No.~C-1509) and NSF (Grant No.~DMR-0325474) for support.  One of us (A.S.) would like to thank M.~R.~Choudhury for helpful discussions.
\end{acknowledgments}


\begin{thebibliography}{27}
\expandafter\ifx\csname natexlab\endcsname\relax\def\natexlab#1{#1}\fi
\expandafter\ifx\csname bibnamefont\endcsname\relax
  \def\bibnamefont#1{#1}\fi
\expandafter\ifx\csname bibfnamefont\endcsname\relax
  \def\bibfnamefont#1{#1}\fi
\expandafter\ifx\csname citenamefont\endcsname\relax
  \def\citenamefont#1{#1}\fi
\expandafter\ifx\csname url\endcsname\relax
  \def\url#1{\texttt{#1}}\fi
\expandafter\ifx\csname urlprefix\endcsname\relax\def\urlprefix{URL }\fi
\providecommand{\bibinfo}[2]{#2}
\providecommand{\eprint}[2][]{\url{#2}}

\bibitem[{\citenamefont{Dresselhaus et~al.}(2007)\citenamefont{Dresselhaus,
  Dresselhaus, Saito, and Jorio}}]{Dresselhaus08ExcitonReview}
For a review, see, e.g.,
\bibinfo{author}{\bibfnamefont{M.~S.} \bibnamefont{Dresselhaus}},
  \bibinfo{author}{\bibfnamefont{G.}~\bibnamefont{Dresselhaus}},
  \bibinfo{author}{\bibfnamefont{R.}~\bibnamefont{Saito}}, \bibnamefont{and}
  \bibinfo{author}{\bibfnamefont{A.}~\bibnamefont{Jorio}},
  \bibinfo{journal}{Annu. Rev. Phys. Chem.} \textbf{\bibinfo{volume}{58}},
  \bibinfo{pages}{719} (\bibinfo{year}{2007}).

\bibitem[{\citenamefont{Ma et~al.}(2005)\citenamefont{Ma, Valkunas, Dexheimer,
  Bachilo, and Fleming}}]{Ma-PRL-2005}
\bibinfo{author}{\bibfnamefont{Y.-Z.} \bibnamefont{Ma}},
  \bibinfo{author}{\bibfnamefont{L.}~\bibnamefont{Valkunas}},
  \bibinfo{author}{\bibfnamefont{S.~L.} \bibnamefont{Dexheimer}},
  \bibinfo{author}{\bibfnamefont{S.~M.} \bibnamefont{Bachilo}},
  \bibnamefont{and} \bibinfo{author}{\bibfnamefont{G.~R.}
  \bibnamefont{Fleming}}, \bibinfo{journal}{Phys.~Rev.~Lett.}
  \textbf{\bibinfo{volume}{94}}, \bibinfo{pages}{157402}
  (\bibinfo{year}{2005}).

\bibitem[{\citenamefont{Ostojic et~al.}(2005)\citenamefont{Ostojic, Zaric,
  Kono, Moore, Hauge, and Smalley}}]{Ostojic-PRL-2005}
\bibinfo{author}{\bibfnamefont{G.~N.} \bibnamefont{Ostojic}},
  \bibinfo{author}{\bibfnamefont{S.}~\bibnamefont{Zaric}},
  \bibinfo{author}{\bibfnamefont{J.}~\bibnamefont{Kono}},
  \bibinfo{author}{\bibfnamefont{V.~C.} \bibnamefont{Moore}},
  \bibinfo{author}{\bibfnamefont{R.~H.} \bibnamefont{Hauge}}, \bibnamefont{and}
  \bibinfo{author}{\bibfnamefont{R.~E.} \bibnamefont{Smalley}},
  \bibinfo{journal}{Phys.~Rev.~Lett.} \textbf{\bibinfo{volume}{94}},
  \bibinfo{pages}{097401} (\bibinfo{year}{2005}).

\bibitem[{\citenamefont{Murakami and Kono}(2008)}]{Murakami-under-review}
\bibinfo{author}{\bibfnamefont{Y.}~\bibnamefont{Murakami}} \bibnamefont{and}
  \bibinfo{author}{\bibfnamefont{J.}~\bibnamefont{Kono}}, \eprint{arXiv:0804.3190v1}.

\bibitem[{\citenamefont{Sheng et~al.}(2005)\citenamefont{Sheng, Vardeny,
  Dalton, and Baughman}}]{Sheng-PRB-2005}
\bibinfo{author}{\bibfnamefont{C.-X.} \bibnamefont{Sheng}},
  \bibinfo{author}{\bibfnamefont{Z.~V.} \bibnamefont{Vardeny}},
  \bibinfo{author}{\bibfnamefont{A.~B.} \bibnamefont{Dalton}},
  \bibnamefont{and} \bibinfo{author}{\bibfnamefont{R.~H.}
  \bibnamefont{Baughman}}, \bibinfo{journal}{Phys.~Rev.~B}
  \textbf{\bibinfo{volume}{71}}, \bibinfo{pages}{125427}
  (\bibinfo{year}{2005}).

\bibitem[{\citenamefont{Russo et~al.}(2006)\citenamefont{Russo, Mele, Kane,
  Rubtsov, Therien, and Luzzi}}]{Russo-PRB-2006}
\bibinfo{author}{\bibfnamefont{R.~M.} \bibnamefont{Russo}},
  \bibinfo{author}{\bibfnamefont{E.~J.} \bibnamefont{Mele}},
  \bibinfo{author}{\bibfnamefont{C.~L.} \bibnamefont{Kane}},
  \bibinfo{author}{\bibfnamefont{I.~V.} \bibnamefont{Rubtsov}},
  \bibinfo{author}{\bibfnamefont{M.~J.} \bibnamefont{Therien}},
  \bibnamefont{and} \bibinfo{author}{\bibfnamefont{D.~E.} \bibnamefont{Luzzi}},
  \bibinfo{journal}{Phys.~Rev.~B} \textbf{\bibinfo{volume}{74}},
  \bibinfo{pages}{041405(R)} (\bibinfo{year}{2006}).

\bibitem[{\citenamefont{Cognet et~al.}(2007)\citenamefont{Cognet, Tsyboulski,
  Rocha, Donyle, Tour, and Weisman}}]{Cognet-Science-2007}
\bibinfo{author}{\bibfnamefont{L.}~\bibnamefont{Cognet}},
  \bibinfo{author}{\bibfnamefont{D.~A.} \bibnamefont{Tsyboulski}},
  \bibinfo{author}{\bibfnamefont{J.~R.} \bibnamefont{Rocha}},
  \bibinfo{author}{\bibfnamefont{C.~D.} \bibnamefont{Donyle}},
  \bibinfo{author}{\bibfnamefont{J.~M.} \bibnamefont{Tour}}, \bibnamefont{and}
  \bibinfo{author}{\bibfnamefont{R.~B.} \bibnamefont{Weisman}},
  \bibinfo{journal}{Science} \textbf{\bibinfo{volume}{316}},
  \bibinfo{pages}{1465} (\bibinfo{year}{2007}).

\bibitem[{\citenamefont{Ben-Avraham and Havlin}(2000)}]{BookBenAvraham}
\bibinfo{author}{\bibfnamefont{D.}~\bibnamefont{ben-Avraham}} \bibnamefont{and}
  \bibinfo{author}{\bibfnamefont{S.}~\bibnamefont{Havlin}},
  \emph{\bibinfo{title}{Diffusion and Reactions in Fractals and Disordered
  Systems}} (\bibinfo{publisher}{Cambridge University Press, Cambridge},
  \bibinfo{year}{2000}).

\bibitem[{\citenamefont{Cardy}(1996)}]{CardyMathBeauty}
\bibinfo{author}{\bibfnamefont{J.~L.} \bibnamefont{Cardy}},
  \emph{\bibinfo{title}{The Mathematical Beauty of Physics}}
  (\bibinfo{publisher}{World Scientific Publishing Co., Singapore},
  \bibinfo{year}{1996}).

\bibitem[{\citenamefont{Marro and Dickman}(1999)}]{BookNonEqlbTransMarro}
\bibinfo{author}{\bibfnamefont{J.}~\bibnamefont{Marro}} \bibnamefont{and}
  \bibinfo{author}{\bibfnamefont{R.}~\bibnamefont{Dickman}},
  \emph{\bibinfo{title}{Nonequilibrium Phase Transitions in Lattice Models}}
  (\bibinfo{publisher}{Cambridge University Press, Cambridge},
  \bibinfo{year}{1999}).

\bibitem[{\citenamefont{Odor}(2004)}]{Odor04RMP}
\bibinfo{author}{\bibfnamefont{G.}~\bibnamefont{Odor}}, \bibinfo{journal}{Rev.
  Mod. Phys.} \textbf{\bibinfo{volume}{76}}, \bibinfo{pages}{663}
  (\bibinfo{year}{2004}).

\bibitem[{\citenamefont{Okubo}(1980)}]{BookOkubo}
\bibinfo{author}{\bibfnamefont{A.}~\bibnamefont{Okubo}},
  \emph{\bibinfo{title}{Diffusion and Ecological Problems: Mathematical
  Models}} (\bibinfo{publisher}{Springer-Verlag, New York},
  \bibinfo{year}{1980}).

\bibitem[{\citenamefont{Hinrichsen}(2000)}]{Hinrichsen00AdvPhys}
\bibinfo{author}{\bibfnamefont{H.}~\bibnamefont{Hinrichsen}},
  \bibinfo{journal}{Adv. Phys.} \textbf{\bibinfo{volume}{49}},
  \bibinfo{pages}{815} (\bibinfo{year}{2000}).

\bibitem[{\citenamefont{Kamenev}(2002)}]{KamanevManybody1}
\bibinfo{author}{\bibfnamefont{A.}~\bibnamefont{Kamenev}},
  \emph{\bibinfo{title}{Strongly Correlated Fermions and Bosons in
  Low-Dimensional Disordered Systems}} (\bibinfo{publisher}{Kluwer Academic
  Press, Dordrecht}, \bibinfo{year}{2002}).

\bibitem[{\citenamefont{Kamenev}(2005)}]{BookKamenevManybody2}
\bibinfo{author}{\bibfnamefont{A.}~\bibnamefont{Kamenev}},
  \emph{\bibinfo{title}{Nanophysics: Coherence and Transport}}
  (\bibinfo{publisher}{Elsevier, Amsterdam}, \bibinfo{year}{2005}).

\bibitem[{\citenamefont{Polya}(1921)}]{PolyaResult}
\bibinfo{author}{\bibfnamefont{G.}~\bibnamefont{Polya}},
  \bibinfo{journal}{Math. Ann.} \textbf{\bibinfo{volume}{84}},
  \bibinfo{pages}{129} (\bibinfo{year}{1921}).

\bibitem[{\citenamefont{Cardy and Tauber}(1998)}]{CardyTauber98JStatPhys}
\bibinfo{author}{\bibfnamefont{J.}~\bibnamefont{Cardy}} \bibnamefont{and}
  \bibinfo{author}{\bibfnamefont{U.}~\bibnamefont{Tauber}},
  \bibinfo{journal}{J. Stat. Phys.} \textbf{\bibinfo{volume}{90}},
  \bibinfo{pages}{1} (\bibinfo{year}{1998}).

\bibitem[{\citenamefont{Torney and McConnell}(1983)}]{Torney-JPC-1983}
\bibinfo{author}{\bibfnamefont{D.~C.} \bibnamefont{Torney}} \bibnamefont{and}
  \bibinfo{author}{\bibfnamefont{H.~M.} \bibnamefont{McConnell}},
  \bibinfo{journal}{J.~Phys.~Chem.} \textbf{\bibinfo{volume}{87}},
  \bibinfo{pages}{1941} (\bibinfo{year}{1983}).

\bibitem[{\citenamefont{Toussaint and Wilczek}(1983)}]{Toussaint-JCP-1983}
\bibinfo{author}{\bibfnamefont{D.}~\bibnamefont{Toussaint}} \bibnamefont{and}
  \bibinfo{author}{\bibfnamefont{F.}~\bibnamefont{Wilczek}},
  \bibinfo{journal}{J.~Chem.~Phys.} \textbf{\bibinfo{volume}{78}},
  \bibinfo{pages}{2642} (\bibinfo{year}{1983}).

\bibitem[{\citenamefont{ben Avraham}(1998)}]{ben-Avraham-PRL-1998}
\bibinfo{author}{\bibfnamefont{D.}~\bibnamefont{ben-Avraham}},
  \bibinfo{journal}{Phys.~Rev.~Lett.} \textbf{\bibinfo{volume}{81}},
  \bibinfo{pages}{4756} (\bibinfo{year}{1998}).

\bibitem[{\citenamefont{Spouge}(1988)}]{Spouge88PRL}
\bibinfo{author}{\bibfnamefont{J.~L.}~\bibnamefont{Spouge}},
  \bibinfo{journal}{Phys. Rev. Lett.} \textbf{\bibinfo{volume}{60}},
  \bibinfo{pages}{871} (\bibinfo{year}{1988}).

\bibitem[{\citenamefont{Karatzas and Shreve}(2004)}]{BookShreve}
\bibinfo{author}{\bibfnamefont{I.}~\bibnamefont{Karatzas}} \bibnamefont{and}
  \bibinfo{author}{\bibfnamefont{S.}~\bibnamefont{Shreve}},
  \emph{\bibinfo{title}{Brownian Motion and Stochastic Calculus}}
  (\bibinfo{publisher}{Springer, New York}, \bibinfo{year}{2004}).

\bibitem[{\citenamefont{Citrin}(1992)}]{Citrin92PRL}
\bibinfo{author}{\bibfnamefont{D.~S.} \bibnamefont{Citrin}},
  \bibinfo{journal}{Phys. Rev. Lett.} \textbf{\bibinfo{volume}{69}},
  \bibinfo{pages}{3393} (\bibinfo{year}{1992}).

\bibitem[{\citenamefont{Spataru et~al.}(2005)\citenamefont{Spataru,
  Ismail-Beigi, Capaz, and Louie}}]{SpataruetAl05PRL}
\bibinfo{author}{\bibfnamefont{C.~D.} \bibnamefont{Spataru}},
  \bibinfo{author}{\bibfnamefont{S.}~\bibnamefont{Ismail-Beigi}},
  \bibinfo{author}{\bibfnamefont{R.~B.}~\bibnamefont{Capaz}}, \bibnamefont{and}
  \bibinfo{author}{\bibfnamefont{S.~G.} \bibnamefont{Louie}},
  \bibinfo{journal}{Phys. Rev. Lett.} \textbf{\bibinfo{volume}{95}},
  \bibinfo{pages}{247402} (\bibinfo{year}{2005}).

\bibitem[{\citenamefont{Perebeinos et~al.}(2005)\citenamefont{Perebeinos,
  Tersoff, and Avouris}}]{PerebeinosetAl05NL}
\bibinfo{author}{\bibfnamefont{V.}~\bibnamefont{Perebeinos}},
  \bibinfo{author}{\bibfnamefont{J.}~\bibnamefont{Tersoff}}, \bibnamefont{and}
  \bibinfo{author}{\bibfnamefont{Ph.}~\bibnamefont{Avouris}},
  \bibinfo{journal}{Nano Lett.} \textbf{\bibinfo{volume}{5}},
  \bibinfo{pages}{2495} (\bibinfo{year}{2005}).

\bibitem[{\citenamefont{Srivastava et~al.}(2008)\citenamefont{Srivastava,
  Htoon, Klimov, and Kono}}]{SrivastavaetAl08PRL}
\bibinfo{author}{\bibfnamefont{A.}~\bibnamefont{Srivastava}},
  \bibinfo{author}{\bibfnamefont{H.}~\bibnamefont{Htoon}},
  \bibinfo{author}{\bibfnamefont{V.~I.} \bibnamefont{Klimov}},
  \bibnamefont{and} \bibinfo{author}{\bibfnamefont{J.}~\bibnamefont{Kono}},
  \bibinfo{journal}{Phys. Rev. Lett.} \textbf{\bibinfo{volume}{101}},
  \bibinfo{pages}{087402} (\bibinfo{year}{2008}).

\bibitem[{\citenamefont{Hoegele et~al.}(2007)\citenamefont{Hoegele, Galland,
  Winger, and Imamoglu}}]{HogeleetAl07PRL}
\bibinfo{author}{\bibfnamefont{A.}~\bibnamefont{Hogele}},
  \bibinfo{author}{\bibfnamefont{C.}~\bibnamefont{Galland}},
  \bibinfo{author}{\bibfnamefont{M.}~\bibnamefont{Winger}}, \bibnamefont{and}
  \bibinfo{author}{\bibfnamefont{A.}~\bibnamefont{Imamoglu}},
  \bibinfo{journal}{Phys. Rev. Lett.} \textbf{\bibinfo{volume}{100}},
  \bibinfo{pages}{217401} (\bibinfo{year}{2008}).

\end{thebibliography}

\end{document}